# WACA: A Hierarchical Weighted Clustering Algorithm optimized for Mobile Hybrid Networks

Matthias R. Brust, Adrian Andronache and Steffen Rothkugel
Faculty of Science, Technology and Communication (FSTC)
University of Luxembourg
Luxembourg
{matthias.brust, adrian.andronache, steffen.rothkugel}uni.lu

*Abstract*—Clustering techniques create hierarchal network structures, called clusters, on an otherwise flat network. In a dynamic environment—in terms of node mobility as well as in terms of steadily changing device parameters—the clusterhead election process has to be re-invoked according to a suitable update policy. Cluster re-organization causes additional message exchanges and computational complexity and it execution has to be optimized. Our investigations focus on the problem of minimizing clusterhead re-elections by considering stability criteria. These criteria are based on topological characteristics as well as on device parameters. This paper presents a weighted clustering algorithm optimized to avoid needless clusterhead re-elections for stable clusters in mobile ad-hoc networks. The proposed localized algorithm deals with mobility, but does not require geographical, speed or distances information.

## I. INTRODUCTION

Multi-hop ad-hoc networks are composed of a collection of devices that communicate with each other over a wireless medium [1]. Such a network can be formed spontaneously whenever devices are in transmission range. Joining and leaving of nodes occurs dynamically, particularly when dealing with mobility in ad-hoc networks. Potential applications of such networks can be found in traffic scenarios, environmental observations, ubiquitous Internet access, and in search and rescue scenarios as described in detail in [2].

Ad-hoc networks emphasize flexibility and survivability of the whole system. However, centralized approaches e.g. for group management and information provisioning do not work well in such settings. Moreover, due to frequent topology changes, connectivity of devices cannot be generally guaranteed. In particular, this makes it hard to disseminate information in a reliable way.

We overcome these limitations inherent to pure ad-hoc networks by (a) establishing local groups of communicating devices in a self-organizing manner and (b) introducing dedicated uplinks to a backbone infrastructure. Such uplinks are used for accessing resources available in the Internet. Additionally, they are employed to directly interconnect distant devices, either within a single partition as well as across different partitions. In practice, uplinks are realized for instance using cellular networks, satellites, or via Wi-Fi hotspots [1]. Hence, ad-hoc networks with devices that provide uplinks are called hybrid wireless networks throughout this paper. Note that uplinks normally imply additional costs and obey lower bandwidth, so that the uplink has to be applied cautiously.

In [3] WACA (*Weighted Application Aware Clustering Algorithm*) is introduced that deals with hybrid wireless networks. WACA fosters efficient information dissemination within the ad-hoc neighborhood as well as limits the use of uplinks to the backbone network. The simulation studies conducted in [3] are based on static network topologies.

The contribution of this paper is to study WACA's performance with respect to dynamic environments. As result of this study, we propose the introduction of a so-called king bonus mechanism in order to optimize the clusterhead election process by stabilizing efficient clusterheads.

The remainder of this paper is organized as follows. Section II describes related work. Section III introduces the mobile hybrid network models. Section IV contains a detailed description of WACA including the king bonus mechanism. The simulation studies conducted are discussed in Section V. The paper finishes with a conclusion in Section VI.

## II. RELATED WORK

Clustering algorithms can be based on criteria such as energy level of nodes, their position, degree, speed and direction. Centralized and distributed approaches can be distinguished, as well as probabilistic and deterministic ones. Probably the most crucial point when dealing with clustering is the criterion how to choose the clusterhead. The number of clusterheads strongly influences the communication overhead, latency, inter- and intra-cluster communication design as well as the update policy (i.e. execution of re-organization of clusters).

One of the first and most influential cluster-based protocols is LEACH [4]. It uses a distributed probabilistic approach. Each node elects itself as a clusterhead with a certain probability based on the desired percentage of the clusterheads in the network, and the last round where it served as a clusterhead. Thus, the role of the clusterhead is probabilistically rotated, which enables to save a large amount of energy.

In [5], a centralized clusterhead election algorithm is presented, where the base station assigns the clusterhead roles based on the energy level and the geographical position of the nodes.

In [6], a centralized algorithm based on fuzzy logic is proposed. The nodes are selected as clusterheads by the base station based on their distances to each other, energy level, and the concentration of the nodes in the region.

Chatterjee et al. [7] propose a distributed deterministic clusterhead selection algorithm, namely WCA (Weighted Clustering Algorithm). For reasons that the proposed WACA clustering algorithm is compared to WCA in this paper, WCA is described in more detail here.

WCA obtains 1-hop clusters with one clusterhead. The election of the clusterhead is based on the weight of each node. For this a heuristic weight function is used that uses distances between the neighbors, degree (number of neighbors), speed of neighboring nodes, and battery power of the node as well as weighting factors to calculate the weight. To obtain this information, WCA assumes to be provided with geographical information or relative distances of one node and its surrounding. The WCA update policy is triggered to be invoked by isolated nodes on demand. Special cases are detachment of current clusterhead and attachment to a new clusterhead. The clusterhead continuously sends a message to its neighbors. The neighbors check if the signal strength decreases what implies that the distance to the clusterhead is increasing. In that case, the node informs its current clusterhead that it detaches and chooses the next available clusterhead. If there is no clusterhead available, the election procedure is evoked to create a new cluster. Observe that the continuous message exchange is a principal drawback of that algorithm.

Tan et al. [8] present a distributed clusterhead selection algorithm where each node computes its priority based on its ID, current communication round, energy level and speed. This information is exchanged within the two-hop neighborhood. The nodes with highest priority become clusterheads.

Early approaches as [9] describe an ID-based clusterhead selection algorithm. Each node in the network is assigned a unique ID. The selection process consists in designating locally the device with the lowest ID as clusterhead. No further parameters are used in this approach.

None of the algorithms introduced before gives guarantees on the resulting network structure, e.g. on the number of the resulting clusterheads. Their effectiveness is evaluated by simulation. In this sense, the aforementioned algorithms realize heuristics. Our approach also falls into this category.

III. MOBILE HYBRID NETWORK SYSTEM MODEL

We define a mobile hybrid network to consist of a set of computers connected by wireless network links. Each device is able to use radio technology like for instance Bluetooth or Wi-Fi for wireless communication in the physical proximity. The considered devices are potentially heterogeneous, since each of them might be supplied with different memory capacities, computational power, and available energy. Some of these devices are supposed to be equipped with uplink-capable adapters such as GSM, 3G or satellite. Devices might be static or mobile; sometimes they may fail, disappear from the network (e.g., due to some obstacles). Thus, the neighborhood of a node changes over time.

*A. Ad-hoc Networking*

In the local ad-hoc neighborhood, nodes potentially have different transmission ranges with bidirectional communication links. This appears as a valid consideration, because wireless protocols such as 802.11 MAC layer require links to be bidirectional for unicasts, too. We furthermore abstract away the details of the MAC and network layer. We assume that every node knows its current one-hop neighbors. Geographical positions of the nodes, however, are not used.

*B. Backbone Networking*

As already mentioned, some devices are equipped with adapters like for GSM, 3G or satellite, thus capable of establishing uplinks to a backbone network. The backbone offers information providing services, possibly organized in a centralized way that is not of interest for this paper. By virtue of registrations, the backbone is aware of the existence of participating devices with uplink-capabilities.

*C. Injection Communication*

Our communication model follows the injection communication paradigm introduced in [10] and further discussed in [11]. Locally, the ad-hoc network devices elect a clusterhead that is in charge of keeping track of local devices. Clusterheads are chosen according to their weight that is calculated by a heuristic weight function (cf. Section IV). Clusterheads might also act as injection points [11]. Injection points maintain a connection to the backbone network and request information related to the common *interests* shared by the devices in the cluster.

The backbone decides when and where to inject fresh information into the multi-hop ad-hoc network by sending it to the injection point. Updates of information are also injected as long as the injection point keeps the connection to the backbone. The injection point disseminates the injected information over a wireless connection like Wi-Fi or Bluetooth to the interested devices according to some criteria. These connections have the advantage of being free of charge. Moreover, these technologies allow higher bandwidth compared to GSM or 3G cellular connections. In a scenario where a cellular flat rate is not available, the injection point is finally also in charge of splitting the cellular connection costs to devices receiving the information. Splitting of costs, however, is an issue of further investigations.

## IV. THE WACA CLUSTERING ALGORITHM

In this section we describe the Weighted Application aware Clustering Algorithm (WACA) and the heuristic weight function used to establish clusters in a self-organizing way. Additionally, the king bonus mechanism is introduced in order to improve the stability of already selected clusterheads, respectively to avoid superfluous re-elections.

### A. Algorithm

One objective of WACA is to avoid network communication overhead during the clusterhead election and clustering process. Therefore, the election of a local clusterhead is based solely on information available locally. To achieve this, each device calculates its own weight based on its device parameters like remaining power, backbone signal strength and topological attributes (cf. Section IV). The weight is recalculated when attribute values change. Each device propagates its own weight as part of the beacon, which is a periodically broadcasted message used in ad-hoc networks to detect devices in communication range. The algorithm only considers direct neighbors, i.e. devices in coverage area of each other.

The pseudo code of the WACA algorithm is shown in Fig. 1. Devices run the algorithm each time the set of neighbor devices changes, e.g. when devices enter or leave the communication range, or when weights are updated. Using the information about the neighborhood, each device elects the neighbor device with the highest

```
Algorithm WACA for device d
Initialization:
1.   w(d) ← calculateWeight(A);
2.   ch(d) ← d;
3.   setBeaconData(w(d), ch(d));
Called when:
Elect clusterhead:
     The set of one-hop neighbor devices of d changes
     The attributes of device d changes
Input:
     A:  {a | a is attribute of devices d}
     N:  {n | n is neighbor of d} // updated neighbors set
     M:  {m | m ∈ N before update(N)} // old neighbors
     W:  {w | w(n) = weight of n, n ∈ N}
     CH: {n | ch(n) is clusterhead of n, n ∈ N}
     k:  The king bonus of device d
Output:
     w(d):         Weight of d
     ch(d):        Clusterhead of d
     isClusterHead: True if ch(d) = d
     isSubHead:    True if there is a n ∈ N, ch(n) = d

01.  wasClusterHead ← isClusterHead;
02.  isClusterHead ← false;
03.  isSubHead ← false;
04.  w(d) ← calculateWeight(A, k);
05.  ch(d) ← d;
06.  for each n ∈ N do
07.      if w(n) > w(d) then
08.          ch(d) ← n;
09.      if ch(n) = d then
10.          isSubHead ← true;
11.  if ch(d) = d then
12.      isClusterHead ← true;
13.  setBeaconData(w(d), ch(d));
14.  if isClusterHead and NOT(wasClusterHead) then
15.      calculateKingBonus(isClusterHead, N, M);
16.  if NOT(isClusterHead) and wasClusterHead then
17.      calculateKingBonus(isClusterHead, N, M);
```

Figure 1. Pseudocode of the WACA algorithm.

weight as clusterhead. Devices that use a mutual clusterhead are called *cluster slaves* of that device.

Clusters are created in a hierarchical fashion. Each device elects exactly one device as its clusterhead, i.e. the neighbor with the highest weight (cf. Fig. 2). This clusterhead also investigates its one-hop neighborhood, similarly electing the device with the highest weight as its clusterhead. This process terminates in case of a device electing itself as its own clusterhead, due to the fact of having the highest weight among all its neighbors. We call all intermediary devices along such clusterhead chains *sub-heads* (Fig. 3). Each device on top of a chain is called a full clusterhead, or, in short, just clusterhead. Hence, in each network partition, multiple clusterheads might coexist.

Only (full) clusterheads might serve as injection points, depending on current interests of the cluster members. Information will be injected to clusterheads only. Sub-heads in turn are responsible for further forwarding the data to other cluster members, i.e. its cluster slaves.

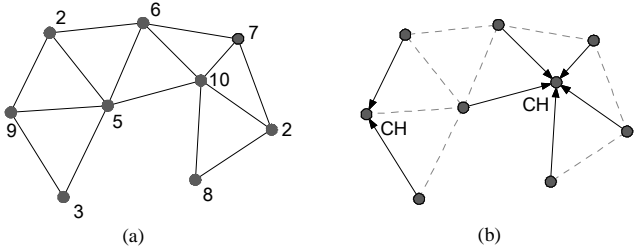

Figure 2. Topology before (a) and after (b) applying WACA. Two clusterheads (CH) are established.

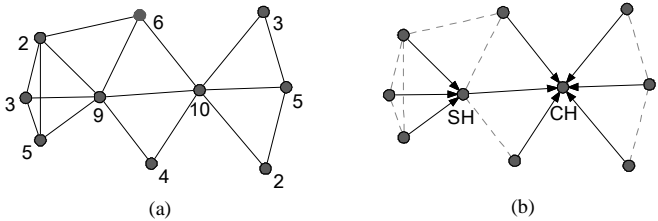

Figure 3. Topology before (a) and after (b) applying WACA. The resulting topology includes one clusterhead (CH) and one sub-head (SH).

As mentioned above, sub-heads are in charge of the same tasks as clusterheads concerning their cluster slaves. The only difference to a clusterhead is the fact that the sub-head will not maintain a backbone connection and will forward the requests and information from its cluster slaves to its own clusterhead. Only in case of a sub-head losing connectivity to its clusterhead it will connect to the backbone itself, thus becoming a clusterhead. Timing issues are of importance in this respect and will be discussed further in Sub-section E.

*B. System Parameters*

Appropriate selection of parameters for calculating weights is a crucial point. Our approach focuses on both, augmenting stability of the clustering topology and fulfilling application requirements. According parameters include device power, signal strength and clustering characteristics.

*1) Device Power*

Nowadays, wireless connections like Wi-Fi or Bluetooth are providing a considerable higher bandwidth than 3G cellular connections. Furthermore, 3G communications also consumes more energy. By definition, clusterheads that also act as injection points rely on both types of connections, both locally as well as for the uplink. They need to keep track of local administration and to provide injected information to interested devices in the own cluster. Hence, the remaining battery power of devices must be considered when electing clusterheads.

*2) Signal Strength for Backbone Connectivity*

In case of cellular networks, devices closer to a 3G CDMA base station perceive higher data rates than devices farther away. A higher distance in turn typically results in intermittent connectivity and lower data rates. Electing devices closer to the base station provides higher data rates for data injection [12]. Once the data has reached the ad-hoc network, it can be disseminated via high bandwidth links, e.g. Wi-Fi, to all devices within that cluster. The influence of the quality of the uplink is much higher than that of the local ad-hoc communication. That is why we explicitly omitted the ad-hoc signal strength from the clusterhead election process.

*3) Dissemination Degree*

A clusterhead is a pivot entity for information distribution. Information requests, cluster management, backbone-driven information injection and also distribution of that information to interested devices are among the tasks of clusterheads. In order to disseminate injected information efficiently, clusterheads should have a higher degree than their neighbors, i.e. more connections to other devices than each device in the communication range. In practice the ideal degree of a device depends strongly on the technology employed, as pointed out in [13]. For instance, the master-slave model used in *Bluetooth* handles up to seven slaves. In that case, a higher degree than seven causes latency in information delivery. Like in [13], we introduce a parameter $dd_I$ that represents the ideal degree for a clusterhead to achieve high throughput.

*4) Local Clustering*

Due to the introduction of sub-heads, WACA creates multi-hop clusters of unknown size. Applying WACA in large network partitions might result in a high number of hops to the clusterhead. This is due to the possibly long chain of sub-heads attached to a clusterhead. To tackle this problem locally, we propose the use of a local clustering coefficient as described in [14] to support well-structured clusters. Informally speaking, the local clustering coefficient $c_L$ is capturing the connectivity between the neighbors of one node. This is described in more detail in the subsequent sub-section. Additionally, the local clustering coefficient can help in identifying topologically unfavorable devices. For instance devices close to partition borders can be assumed to leave the partition earlier than more centralized ones. The local clustering coefficient is a local aid to limit the cluster size fostering scalability of WACA.

*C. Heuristic Weight Function*

The system parameters are combined with appropriate weighing factors. The flexibility of changing the weighing factors allows applying the algorithms for very different networks as well as applications. Due to the fact that the network topology is built based on the weight of the devices, the weight calculation plays a central role in our algorithm. In order to determine the weight, following calculations have to be performed. Hereby we assume that the neighbor discovery service already filled the neighbor list $N(d)$ on one device $d$ with the IDs of those devices within the transmission range of $d$. $D$ represents the set of devices.

$$N(d) = \bigcup_{d' \in D, d' \neq d} \{dist(d,d') < r\}. \tag{1}$$

**Device power.** Given a device $d$ with available power $P(d)$, then calculate the power-appropriateness $P_A$ of device $d$ as

$$P_A = \frac{3}{2} + \frac{1}{2}\log\left(P(d) - \frac{3}{5}\right). \tag{2}$$

**Signal strength.** Usually the strength of the backbone network signal is available on each device, e.g. the signal strength to a cellular network base station. Let $s$ be the strength of the signal, given by a value between 0 and 1.

**Dissemination degree.** Compute the difference between ideal degree $dd_I$ for device $d$ and real degree $|N(d)|$ (cf. [13]) as

$$\Delta dd = 1 - \frac{|N(d) - dd_I|}{dd_I}. \tag{3}$$

**Local clustering coefficient.** Compute the local clustering coefficient $c_L$ of one device $d$ that is defined as

$$c_L = \frac{|N(d)|}{n(n-1)/2}, \tag{4}$$

where $|N(d)|$ is the number of links in the neighborhood of $d$ and $n(n-1)/2$ is the number of all possible links, whereby $n$ is the number of all devices.

**Calculating weight function.** Calculate the total weight of a device $d$ as

$$W_d = wf_1 P_A + wf_2 s + wf_3 c_L + wf_4 \Delta dd, \tag{5}$$

where $wf_1$, $wf_2$, $wf_3$, and $wf_4$ are weighing factors choosing according requirements.

Choosing a clusterhead depends on the parameters described and the related weighing factors. Due to the different performance of the mobile device batteries the energy level of the batteries cannot be used as metric for $P$. For instance a notebook with 70% battery level will mostly outlive in the injection point role a PDA that has a battery level of 100%. Thus, parameter $P_A$ represents the power-appropriateness being a clusterhead for the task at hand, e.g. receiving a video clip of a certain size. Albeit it is an asset to have more remaining power than necessary to perform the task at hand, that additional power has less influence on the calculation. For this reason, we choose a log-function for calculating the device power parameter. A power-appropriate device may suffer from low bandwidth due to bad connectivity with the base station. The current signal strength is represented as parameter $s$ as described above. Taking the number of neighbor devices, it is reasonable to assume that there is a range of values most appropriate for the MAC layer. Bluetooth piconets and its limitation to at most seven active slaves is one concrete example for that. The parameter $\Delta dd$ reflects the deviation of the number of neighbors in a current setting from that ideal. The local clustering coefficient aims at balancing the cluster structure in terms of selecting appropriate clusterheads. The calculation results in preferring devices topologically located in the center of a device group. As a side-effect, the impact of device mobility is reduced [15].

### D. The King Bonus Mechanism

The king bonus mechanism aims at avoiding superfluous clusterhead elections in mobile networks, thus fostering the stability of the cluster topology. The mechanism is designed for mobile environments where relative device mobility within a group is low, or, in other words, devices forming groups sticking together. In this scenario, a device with higher weight as the clusterhead passing or crossing the cluster will be elected as clusterhead. Such devices will soon disappear from the neighborhood after having passed the cluster. Thus, it would be beneficial *not* to re-organize the cluster temporarily only, even if the new device is of considerably higher weight.

To avoid the superfluous election of crossing devices, WACA is increasing the weight of the clusterheads which have a stable neighborhood over time. The value that is added to the weight of the clusterheads is called *king bonus*. Devices without neighbors—which automatically elect themselves as clusterhead—will not increase their weight. If a clusterhead gets new neighbors or loses cluster members, the neighborhood will be considered unstable and the king bonus will be reduced adequately. In this work we used following formula for the king bonus calculation of a device $d$ where $N$ is the updated set of neighbor devices and $M$ is the old set of the neighbor devices:

$$k(d) = \begin{cases} 0, \text{ if device } d \text{ is not clusterhead} \\ k(d)+33, \text{ if } k(d)<99 \\ k(d)-\left(k(d)\cdot\frac{|M\setminus N|+|N\setminus M|}{|N|+|M|}\right), \text{ if } N \neq M \end{cases}$$

The pseudo code of the king bonus algorithm can be seen in the method *calculateKingBonus* (cf. Figure 4). WACA invokes the method *calculateKingBonus* if (a) the own device is elected as clusterhead, (b) the own device has lost the cluster leadership or (c) the own device is clusterhead and the set of neighbor devices has

**Method** $calculateKingBonus$
**Called when :**
    $d$ gets the clusterhead state
    $d$ loses the clusterhead state
    $S$et of one-hop neighbor devices of $d$ changes ($N != M$)
**Input :**
    $isClusterHead : True$ **if** $d$ is clusterhead
    $N : \{n \mid n \text{ is neighbor of } d\}$ // $updated\ neighbors\ set$
    $M : \{m \mid m \in N \text{ before } update(N)\}$ // $old\ neighbors\ set$
**Output :**
    $k : A$ integer value between 0 and 99

01. **if** $NOT(isClusterHead)$ **then**
02.     $k \leftarrow 0$
03.     $w(d) = calculateWeight(A, k)$
04.     **return**;
05. **if** $\#N = 0$ **then**
06.     $k \leftarrow 0$
07.     $w(d) = calculateWeight(A, k)$
08.     **return**;
09. **if** $k < 99$ **then**
10.     $s \leftarrow 0;$
11.     **if** $N != M$ **then**
12.         $s \leftarrow calculateStabilityCoefficient(N, M)$
13.         $k \leftarrow k - (k * s)$
14.     **else**
14.         $k \leftarrow k + 33$
15.     $w(d) \leftarrow calculateWeight(A, k)$
16.     wait 3 seconds **then**
17.         $calculateKingBonus(isClusterHead, N, M)$

**Method** $calculateStabilityCoefficient$
**Called when :**
    *The updated neighbor set N is different from the old neighbor set M*
**Input :**
    $N : \{n \mid n \text{ is neighbor of } d\}$ // $updated\ neighbors\ set$
    $M : \{m \mid m \in N \text{ before } update(N)\}$ // $old\ neighbors\ set$
**Output :**
    $s : $ A double value between 0 and 1
01. $lostNeighbors = |M \setminus N|$
02. $newNeighbors = |N \setminus M|$
03. $s = (lostNeighbors + newNeighbors) / (|N| + |M|)$
04. **return** $s$

Figure 4. Pseudocode for calculating king bonus and stability coefficient.

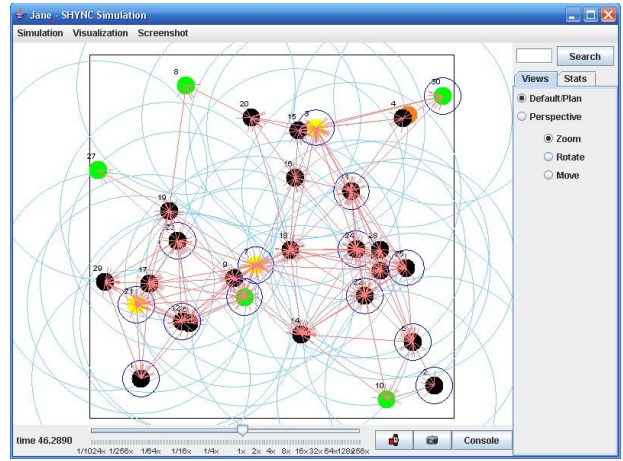

Figure 5. The simulation environment running WACA. Clusterheads and subheads are filled in green and yellow respectively.

changed. The king bonus algorithm will check first if the own device is still clusterhead (pseudo code, line 01). If not, then the king bonus will simply be set to 0 (line 02-04). If the own device is clusterhead but it has no neighbors, then the king bonus will also be set to 0 (line 06-08).

If the king bonus didn't reach the maximum (line 07), the algorithm will check if the neighborhood has changed (line 09) and adequately reduce the king bonus (line 12-13). If the neighborhood didn't change, the king bonus will be increased by 33 (line 15) each 3 seconds until reaching 99. Each time the value of the king bonus changes, the weight of the device will be re-calculated (line 03, 07, 16).

*E. Update Policy and Message Complexity*

When the ad-hoc network is initially established, each device calculates the own weight and disseminates it through beaconing in the neighborhood as described in Section IV. After receiving the beacon from all neighbors, a device will elect the neighbor with the highest weight as its clusterhead. During the lifetime of the ad-hoc network, re-election of clusterheads can occur, due to possible changes of parameters in the neighborhood. The devices are tracking changes of the own parameters, recalculate the own weight and update the beacon. Thus, the devices are always up to date with respect to the current weight of their neighbors.

In case the clusterhead vanishes, e.g. moves out of the cluster's communication range or is switched off, each cluster slave will check its neighbor list and elect the one with the highest weight as new clusterhead. If the system parameters of the clusterhead change so that the device cannot accomplish the injection point tasks anymore, then the weight will be drastically lowered, thus inducing a re-election in the neighborhood.

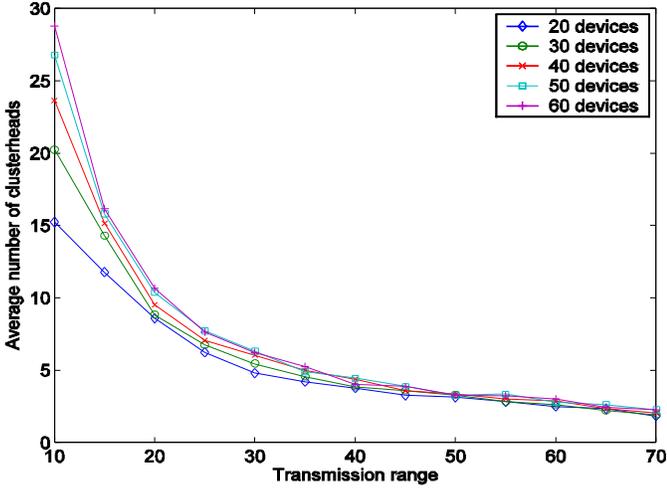

Figure 6.  Average number of clusterheads.

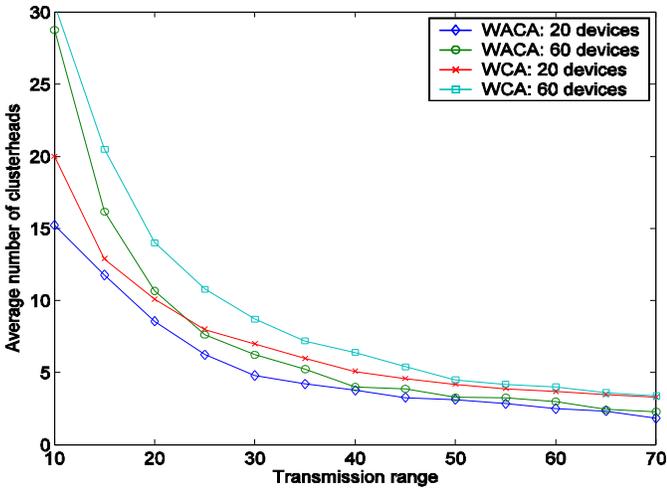

Figure 7.  Comparing the number of clusterheads of WACA and WCA.

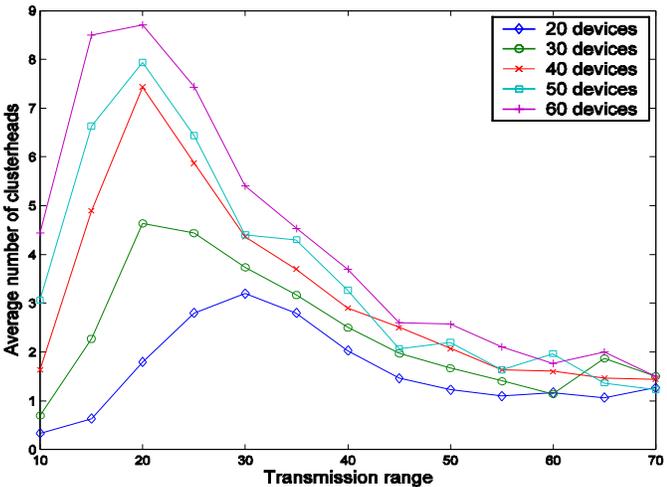

Figure 8.  The average number of sub-heads for different number of devices using WACA.

The WACA algorithm calculates a total weight on each device. After exchanging these weights to the neighbors the device with the highest weight is considered to be a clusterhead. Since each device has to send the weight to its neighbors, the message complexity is $O(n)$, where $n$ is the number of mobile devices in the one-hop network neighborhood.

## V. SIMULATION STUDY

We choose a 100×100 unit square as basic simulation setting. A number of $N$ nodes are deployed uniformly by random using a validated random number generator initialized by independent seeds. In the simulation the number was set to values between 20 and 60. The transmission range varied between 10 and 70 with a fixed step of 5. The weighing values were set to $wf_1 = 0.9$, $wf_2 = 1$, $wf_3 = 0.85$, $wf_4 = 0.65$. Observe that for different application requirements, the weighing factors have to be adjusted (cf. Figure 5). All results are averaged over 30 simulation runs.

### A. Static Network

The first experiment provides a detailed report of the average number of clusterheads or clusters, respectively, for the different values of $N$ (Figure 6). For all values chosen for $N$ the average number of clusterheads decreases when increasing the transmission range. We argue that this asymptotic behavior results from clusterheads with a larger sending range covering an exponentially larger area.

In a second experiment we compared the average number of clusters of WACA with that of WCA [13]. For this, $N$ is set to 20 and 60 using both algorithms. The transmission range is varied as described above. The results show that the average number of clusterheads using WACA in both cases is below of that of WCA (Figure 7). Not illustrated, but shown through further simulation, this also holds for the cases where $N = 30, 40, 50$. As mentioned above, reducing the number of clusterheads strongly influences the communication overhead, latency, inter- and intra-cluster communication design as well as execution of re-organization of clusters. We correlate this performance improvement due to the fact that sub-heads have been introduced to allow multi-hop clusters, but keeping the clusters well-formed.

Further, we investigated the number of sub-heads for the same setting of the first experiment. Results are reported in detail in Fig. 8. The number of sub-heads increases as the transmission range increases, and reaches a peak when transmission range is between 20 and 30. Further increase of the transmission range results in a decrease of the average number of sub-heads.

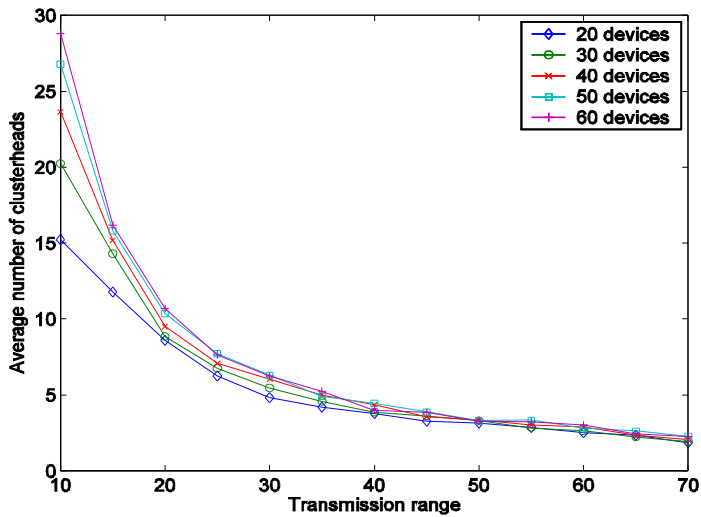

Figure 9. Average number of clusterhead reaffiliations.

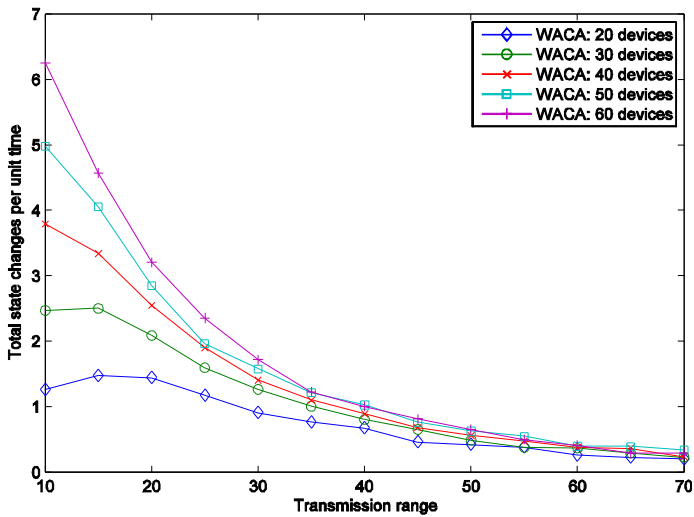

Figure 10. Total number of node state change.

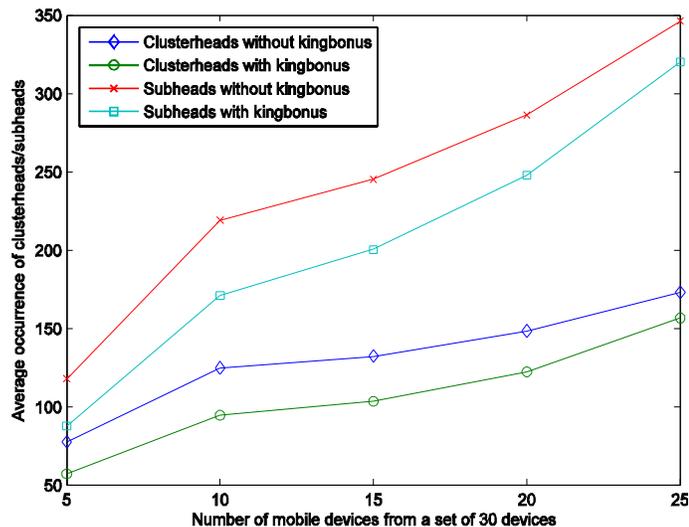

Figure 11. Average number of clusterheads and subhead with and without king bonus mechanism.

This behavior can be explained by the fact that sub-heads cannot be easily established in cases where the transmission range is very low, because clusters tend to be one-hop structures. When increasing the transmission range clusters are getting bigger, encompassing more devices, and an increasing number of sub-heads are established. Further increasing the transmission range results in fewer sub-heads, because the clusterhead can reach more devices directly without the need for an intermediary sub-head. This explains the asymptotic decrease of the average number of sub-heads.

*B. Mobile Networks*

In this simulation set we introduce mobility to our settings. The random waypoint model was used in order to be able to compare the results to WCA. For this, the velocity of nodes was set to 5 units per seconds. The simulation time was set to 900 seconds. As in the previous experiments we conducted the experiments with 20, 30, 40, 50, and 60 devices with transmission range 10 to 70 using an increment of 5.

In Figure 9 the average number of clusterhead re-affiliations per second (i.e. how often all devices have changed their clusterhead) were measured. Correspondingly to the experiment in the static network, Figure 9 shows that the number of clusterhead re-affiliations per second is strongly decreasing when increasing the transmission range. Until transmission range 30 that number is dependent from the device number, while they are almost similar for transmission ranges higher than 30. The density of a network can be augmented by increasing either the transmission range or the number of devices. Figure 9 shows furthermore that above transmission range of 30 units, there is no significant change in the number of re-affiliations when the number of devices or the transmission range is increased further.

Comparing to WCA, the number of clusterhead re-affiliations are higher for transmission ranges between 10 and 20, but are considerable lower for transmission ranges between 20 and 70. Furthermore in WCA the number of re-affiliations is more sensible to device density than WACA as Figure 9 illustrates.

In Figure 10 the total numbers of node state changes are shown. The nodes can change their state between the states clusterhead, subhead or slave. The number of state changes is significantly higher for transmission ranges between 10 and 30 than for transmission ranges between 30 and 70.

The king bonus aims at stabilizing efficient clusterheads avoiding superfluous re-elections and state changes. The experiment results for stability of clusterheads and subheads respectively are shown in

Figure 11. For these experiments we used a set of 30 devices with a transmission range of 30 units. The x-axis shows the number of moving devices while the remaining ones are assumed to be stationary. Figure 11 shows that the king bonus mechanism drastically reduces the total number of clusterheads or subheads respectively. This positive effect of the king bonus mechanism is slightly decreasing by increasing the total number of mobile devices.

## VI. CONCLUSION

We introduced a hierarchical weighted clustering algorithm that attempts to take device properties as well as application requirements into account for clustering by using a heuristic weight function. This way, WACA can be explicitly fine-tuned for different application demands. WACA is explicitly designed for hybrid networks, i.e. the symbiotic combination of multiple ad-hoc network partitions inter-linked by a backbone network.

Results have shown that the average number of clusterheads can be decreased using WACA compared to the WCA clustering algorithm. Note that in contrary to WCA, WACA creates multi-level hierarchical clusters. Additionally, the WACA algorithm does not depend on geographic positions, device speed or on distances between neighbors, which are hard to determine. This makes the implementation of the algorithm on real devices more suitable. WACA works on local information only and supports well-formed multi-hop clusters, realized by introducing cluster sub-heads.

Further results show the performance of WACA in mobile environments. We conclude that WACA minimizes clusterhead re-election in comparison to WCA. Further minimization was done in mobile networks by introducing the king bonus mechanism that avoids clusterhead elections of devices that are just crossing stable clusters.

## REFERENCES

[1] O. Dousse, P. Thiran, and M. Hasler, "Connectivity in Ad hoc and Hybrid Networks," in *INFOCOM'02*, 2002.
[2] P. Santi, *Topology Control in Wireless Ad Hoc and Sensor Networks*: Wiley, 2005.
[3] A. Andronache, M. R. Brust, and S. Rothkugel, "Multimedia Content Distribution in Hybrid Wireless using Weighted Clustering," in *2nd ACM Workshop on Wireless Multimedia Networking and Performance Modeling*, Malaga, Spain, 2006.
[4] W. R. Heinzelman, A. Chandrakasan, and H. Balakrishnan, "Energy-efficient communication protocol for wireless microsensor networks," *System Sciences, 2000. Proceedings of the 33rd Annual Hawaii International Conference on,* p. 10, 2000.
[5] H. G. Luo, F. G. Ye, J. G. Cheng, S. G. Lu, and L. G. Zhang, "TTDD: Two-Tier Data Dissemination in Large-Scale Wireless Sensor Networks," *Wireless Networks,* vol. 11, pp. 161-175, 2005.
[6] I. Gupta, D. Riordan, and S. Sampalli, "Cluster-head election using fuzzy logic for wireless sensor networks," *Communication Networks and Services Research Conference, 2005. Proceedings of the 3rd Annual,* pp. 255-260, 2005.
[7] M. Chatterjee, S. K. Das, and D. Turgut, "A Weight Based Distributed Clustering Algorithm for Mobile ad hoc Networks," *Proceedings of the 7th International Conference on High Performance Computing, LNCS 1970,* pp. 511-524, 2000.
[8] H. Tan, W. Zeng, and L. Bao, "PATM: Priority-Based Adaptive Topology Management for Efficient Routing in Ad Hoc Networks," in *International Conference on Computational Science*, 2005, pp. 485-492.
[9] A. Ephremides, J. E. Wieselthier, and D. J. Baker, "A design concept for reliable mobile radio networks with frequency hopping signaling," *Proceedings of the IEEE,* vol. 75, pp. 56-73, 1987.
[10] M. R. Brust and S. Rothkugel, "A Communication Model for Adaptive Service Provisioning in Hybrid Wireless Networks," in *3rd WSEAS International Conference on Information Security, Hardware/Software Co-design and Computer Networks (ISCOCO 2004)*, 2004.
[11] S. Rothkugel, M. R. Brust, and C. H. C. Ribeiro, "Inquiring the Potential of Evoking Small-World Properties for Self-Organizing Communication Networks," in *5th International Conference on Networking (ICN 06)*, Mauritius, 2006.
[12] H. Luo, R. Ramjee, P. Sinha, L. E. Li, and S. Lu, "UCAN: a unified cellular and ad-hoc network architecture," *Proceedings of the 9th annual international conference on Mobile computing and networking,* pp. 353-367, 2003.
[13] M. C. M. Chatterjee, S. Das, and D. C. M. Turgut, "WCA: A Weighted Clustering Algorithm for Mobile Ad Hoc Networks," *Cluster Computing,* vol. 5, pp. 193-204, 2002.
[14] D. J. Watts and H. Strogatz, "Collective Dynamics of 'Small World' Networks," *Nature,* vol. 393, pp. 440-442, 1998.
[15] M. R. Brust, A. Andronache, S. Rothkugel, and Z. Benenson, "Topology-based Clusterhead Candidate Selection in Wireless Ad-hoc and Sensor Networks," in *2nd IEEE/ACM International Workshop on Software for Sensor Networks (SensorWare 2007)*, Bangalore, India, 2007.